\documentclass[conference, a4paper]{IEEEtran}
\IEEEsettopmargin{t}{1.92cm}

\IEEEoverridecommandlockouts
\usepackage{cite}
\usepackage{amsmath,amssymb,amsfonts}
\usepackage{algorithmic}
\usepackage{graphicx}
\usepackage{textcomp}
\usepackage{xcolor}
\usepackage{hyperref}

\usepackage{tikz}
\usepackage{pgfplots}
\usetikzlibrary{shapes,arrows,positioning,calc, matrix,spy}
\usepackage[nolist]{acronym}
\usetikzlibrary{external}
\newcommand{\round}[1]{\ensuremath{\lfloor#1\rceil}}
\usepackage{booktabs}
\usepackage{xfrac}

\let\svthefootnote\thefootnote
\newcommand\blankfootnote[1]{%
  \let\thefootnote\relax\footnotetext{#1}%
  \let\thefootnote\svthefootnote%
}
\let\svfootnote\footnote
\renewcommand\footnote[2][?]{%
  \if\relax#1\relax%
    \blankfootnote{#2}%
  \else%
    \if?#1\svfootnote{#2}\else\svfootnote[#1]{#2}\fi%
  \fi
}

\newcommand*{\figures}{graphics}

\definecolor{cb-1}{HTML}{4477AA}
\definecolor{cb-2}{HTML}{EE6677}
\definecolor{cb-3}{HTML}{228833}
\definecolor{cb-4}{HTML}{CCBB44}
\definecolor{cb-5}{HTML}{66CCEE}
\definecolor{cb-6}{HTML}{AA3377}
\definecolor{cb-7}{HTML}{BBBBBB}

\pgfplotscreateplotcyclelist{cb list}{
	cb-1,cb-2,cb-3,cb-4,cb-5,cb-6,cb-7
}

\definecolor{kit-green100}{rgb}{0,.59,.51}
\definecolor{kit-green70}{rgb}{.3,.71,.65}
\definecolor{kit-green50}{rgb}{.50,.79,.75}
\definecolor{kit-green30}{rgb}{.69,.87,.85}
\definecolor{kit-green15}{rgb}{.85,.93,.93}
\definecolor{KITgreen}{rgb}{0,.59,.51}

\definecolor{KITpalegreen}{RGB}{130,190,60}
\colorlet{kit-maigreen100}{KITpalegreen}
\colorlet{kit-maigreen70}{KITpalegreen!70}
\colorlet{kit-maigreen50}{KITpalegreen!50}
\colorlet{kit-maigreen30}{KITpalegreen!30}
\colorlet{kit-maigreen15}{KITpalegreen!15}

\definecolor{KITblue}{rgb}{.27,.39,.66}
\definecolor{kit-blue100}{rgb}{.27,.39,.67}
\definecolor{kit-blue70}{rgb}{.49,.57,.76}
\definecolor{kit-blue50}{rgb}{.64,.69,.83}
\definecolor{kit-blue30}{rgb}{.78,.82,.9}
\definecolor{kit-blue15}{rgb}{.89,.91,.95}

\definecolor{KITyellow}{rgb}{.98,.89,0}
\definecolor{kit-yellow100}{cmyk}{0,.05,1,0}
\definecolor{kit-yellow70}{cmyk}{0,.035,.7,0}
\definecolor{kit-yellow50}{cmyk}{0,.025,.5,0}
\definecolor{kit-yellow30}{cmyk}{0,.015,.3,0}
\definecolor{kit-yellow15}{cmyk}{0,.0075,.15,0}

\definecolor{KITorange}{rgb}{.87,.60,.10}
\definecolor{kit-orange100}{cmyk}{0,.45,1,0}
\definecolor{kit-orange70}{cmyk}{0,.315,.7,0}
\definecolor{kit-orange50}{cmyk}{0,.225,.5,0}
\definecolor{kit-orange30}{cmyk}{0,.135,.3,0}
\definecolor{kit-orange15}{cmyk}{0,.0675,.15,0}

\definecolor{KITred}{rgb}{.63,.13,.13}
\definecolor{kit-red100}{cmyk}{.25,1,1,0}
\definecolor{kit-red70}{cmyk}{.175,.7,.7,0}
\definecolor{kit-red50}{cmyk}{.125,.5,.5,0}
\definecolor{kit-red30}{cmyk}{.075,.3,.3,0}
\definecolor{kit-red15}{cmyk}{.0375,.15,.15,0}

\definecolor{KITpurple}{RGB}{160,0,120}
\colorlet{kit-purple100}{KITpurple}
\colorlet{kit-purple70}{KITpurple!70}
\colorlet{kit-purple50}{KITpurple!50}
\colorlet{kit-purple30}{KITpurple!30}
\colorlet{kit-purple15}{KITpurple!15}

\definecolor{KITcyanblue}{RGB}{80,170,230}
\colorlet{kit-cyanblue100}{KITcyanblue}
\colorlet{kit-cyanblue70}{KITcyanblue!70}
\colorlet{kit-cyanblue50}{KITcyanblue!50}
\colorlet{kit-cyanblue30}{KITcyanblue!30}
\colorlet{kit-cyanblue15}{KITcyanblue!15}
\let\j\relax
\newcommand{\j}{\mathrm{j}}

\newcommand*{\vect}[1]{\boldsymbol{#1}}
\newcommand*{\mat}[1]{\MakeUppercase{\boldsymbol{#1}}}

\def\BibTeX{{\rm B\kern-.05em{\sc i\kern-.025em b}\kern-.08em
    T\kern-.1667em\lower.7ex\hbox{E}\kern-.125emX}}
\begin{document}
\begin{acronym}[TROLOLO]
  \acro{ACF}{auto correlation function}
  \acro{ADC}{analog to digital converter}
  \acro{AE}{autoencoder}
  \acro{ASK}{amplitude shift keying}
  \acro{AoA}{angle of arrival}
  \acro{AWGN}{additive white Gaussian noise}
  \acro{BER}{bit error rate}
  \acro{BCE}{binary cross entropy}
  \acro{BMI}{bit-wise mutual information}
  \acro{BPSK}{binary phase shift keying}
  \acro{BP}{backpropagation}
  \acro{BSC}{binary symmetric channel}
  \acro{CAZAC}{constant amplitude zero autocorrelation waveform}
  \acro{CDF}{cumulative distribution function}
  \acro{CE}{cross entropy}
  \acro{CNN}{concolutional neural network}
  \acro{CP}{cyclic prefix}
  \acro{CRB}{Cramér-Rao bound}
  \acro{CRC}{cyclic redundancy check}
  \acro{CSI}{channel state information}
  \acro{DFT}{discrete Fourier transform}
  \acro{DNN}{deep neural network}
  \acro{DOCSIS}{data over cable services}
  \acro{DPSK}{differential phase shift keying}
  \acro{DSL}{digital subscriber line}
  \acro{DSP}{digital signal processing}
  \acro{DTFT}{discrete-time Fourier transform}
  \acro{DVB}{digital video broadcasting}
  \acro{ELU}{exponential linear unit}
  \acro{ESPRIT}{Estimation of Signal Parameter via Rotational Invariance Techniques}
  \acro{FFNN}{feed-forward neural network}
  \acro{FFT}{fast Fourier transform}
  \acro{FIR}{finite impulse response}
  \acro{GD}{gradient descent}
  \acro{GF}{Galois field}
  \acro{GMM}{Gaussian mixture model}
  \acro{GMI}{generalized mutual information}
  \acro{ICI}{inter-channel interference}
  \acro{IDE}{integrated development environment}
  \acro{IDFT}{inverse discrete Fourier transform}
  \acro{IFFT}{inverse fast Fourier transform}
  \acro{IIR}{infinite impulse response}
  \acro{ISI}{inter-symbol interference}
  \acro{JCAS}{joint communication and sensing}
  \acro{KKT}{Karush-Kuhn-Tucker}
  \acro{kldiv}{Kullback-Leibler divergence}
  \acro{LDPC}{low-density parity-check}
  \acro{LLR}{log-likelihood ratio}
  \acro{LTE}{long-term evolution}
  \acro{LTI}{linear time-invariant}
  \acro{LR}{logistic regression}
  \acro{MAC}{multiply-accumulate}
  \acro{MAP}{maximum a posteriori}
  \acro{MLP}{multilayer perceptron}
  \acro{ML}{machine learning}
  \acro{MSE}{mean squared error}
  \acro{MLSE}{maximum-likelihood sequence estimation}
  \acro{NN}{neural network}
  \acro{OFDM}{orthogonal frequency-division multiplexing}
  \acro{OLA}{overlap-add}
  \acro{PAPR}{peak-to-average-power ratio}
  \acro{PDF}{probability density function}
  \acro{pmf}{probability mass function}
  \acro{PSD}{power spectral density}
  \acro{PSK}{phase shift keying}
  \acro{QAM}{quadrature amplitude modulation}
  \acro{QPSK}{quadrature phase shift keying}
  \acro{radar}{radio detection and ranging}
  \acro{RC}{raised cosine}
  \acro{RCS}{radar cross section}
  \acro{RMSE}{root mean squared error}
  \acro{RNN}{recurrent neural network}
  \acro{ROM}{read-only memory}
  \acro{RRC}{root raised cosine}
  \acro{RV}{random variable}
  \acro{SER}{symbol error rate}
  \acro{SNR}{signal-to-noise ratio}
  \acro{SINR}{signal-to-noise-and-interference ratio}
  \acro{SPA}{sum-product algorithm}
  \acro{VCS}{version control system}
  \acro{WLAN}{wireless local area network}
  \acro{WSS}{wide-sense stationary}
\end{acronym}

\title{Autoencoder-based Joint Communication and Sensing of Multiple Targets
	%A Joint Communication and multiple Target Sensing Autoencoder
	%Multiple Target Sensing in a Joint Communication and Sensing Autoencoder
	%A Joint Communication and Sensing Neural Network Autoencoder for Sensing of multiple Targets\\
\thanks{This work has received funding 
	%in part from the European Research Council
	%(ERC) under the European Union’s Horizon 2020 research and innovation
	%programme (grant agreement No. 101001899) and in part 
	from the German
	Federal Ministry of Education and Research (BMBF) within the project
	Open6GHub (grant agreement 16KISK010).}

}
%\author{Charlotte Muth, Laurent Schmalen\\ Karlsruhe Institute of Technology, Germany, \{firstname.lastname\}@kit.edu}

\author{\IEEEauthorblockN{Charlotte Muth and Laurent Schmalen}
	\IEEEauthorblockA{Communications Engineering Lab (CEL), Karlsruhe Institute of Technology (KIT)\\ 
		Hertzstr. 16, 76187 Karlsruhe, Germany, 
		Email: \texttt{\{first.last\}@kit.edu}\vspace*{-1ex}}
}

%\author{\IEEEauthorblockN{1\textsuperscript{st} Charlotte Muth}
%\IEEEauthorblockA{\textit{Communication Engineering Lab} \\
%\textit{Karlsruhe Institute of Technology}\\
%Karlsruhe \\
%\textit{firstname}.\textit{lastname}@kit.edu}
%\and
%\IEEEauthorblockN{2\textsuperscript{nd} Laurent Schmalen}
%\IEEEauthorblockA{\textit{Communication Engineering Lab} \\
%\textit{Karlsruhe Institute of Technology}\\
%Karlsruhe \\
%\textit{firstname}.\textit{lastname}@kit.edu}
%\and
%\and
%\and
%\and

\maketitle

\begin{abstract}
We investigate the potential of \acp{AE} for building a \ac{JCAS} system that enables communication with one user while detecting multiple radar targets and estimating their positions.
Foremost, we develop a suitable encoding scheme for the training of the \ac{AE} and for targeting a fixed false alarm rate of the target detection during training. We compare this encoding with the classification approach using one-hot encoding for radar target detection.
Furthermore, we propose a new training method that complies with possible ambiguities in the target locations. We consider different options for training the detection of multiple targets. We can show that our proposed approach based on permuting and sorting can enhance the angle estimation performance so that single snapshot estimations with a low standard deviation become possible.
We outperform an \acused{ESPRIT}\ac{ESPRIT} benchmark for small numbers of measurement samples.
\end{abstract}


%\section{Introduction}
%This document is a model and instructions for \LaTeX.
%Please observe the conference page limits.

\acresetall
\section{Introduction}
%%% Paragraph 1: Motivation
Electromagnetic sensing and radio communications remain
vital services for society, yet an increase in their sustainability, and consequently in their efficiency, is of rising importance. We can increase spectral and energy efficiency by combining radio communication and sensing into one waveform compared to operating two separate systems. Therefore, this work focuses on the codesign of both functionalities in a \ac{JCAS} system.
So far, standardized approaches for localization and communication, such as the LTE
Positioning Protocol (LPP), or the New Radio Positioning protocol A (NRPPa), need the cooperation of the user equipment to localize it. 
The future 6G network is envisioned to natively support \ac{JCAS} by extending sensing capabilities to non-cooperating targets, such as objects without communication capabilities, and performing general sensing
of the surroundings \cite{Wild2021}. From this approach, we expect to increase spectral efficiency by making spectral resources accessible to communication while maintaining their use for sensing. Simultaneously, we predict an increase in energy efficiency because of the dual-use
of a joint waveform.

%%% Paragraph 2
%% Specific problem considered

%%% Paragraph 4
%% state of the art
In the radar community, the integration of communication capabilities into sensing signals to enhance a standard radar signal with an information sequence for a possible receiver has already been studied \cite{Lampel2019}. %So has the topic of establishing the cooperation of communication and sensing systems.
A well-studied approach to combined communication and sensing is OFDM radar \cite{Sturm2011, Braun2010}. OFDM radar enables the robust detection of objects while maintaining its communication capabilities through careful signal processing.
%However, the OFDM waveform is not robust in highly-mobile scenarios that will particularly profit from sensing capabilities~\cite{Gaudio2020}.
However, there is a growing interest in data-driven approaches based on \ac{ML} since they can overcome deficits that model-based techniques as used in OFDM face. Especially at higher frequencies used for sensing applications, which will become more important in 6G, these deficits become more pronounced because of hardware imperfections \cite{MateosRamos2021}. \ac{ML} is expected to be prevalent in 6G since its use has matured in communication as well as in radar processing \cite{Wild2021}.
\Acp{AE} have been studied for communication systems, e.g., \cite{OShea2017,Cammerer2020}, and in the context of radar
\cite{JaraboAmores2008, Fuchs2020}. 
In \cite{MateosRamos2021}, an \ac{AE} for \ac{JCAS} in a single-carrier system has been proposed and has shown to robustly perform close to a maximum a-posteriori ratio test detector benchmark for single snapshot evaluation and one possible radar target.

%%% Paragraph 3
%% specific topic that this work covers
In this paper, we explore the monostatic sensing capabilities of a wireless single-carrier communication system. We use an \ac{AE} approach and study the influence of multi-target sensing and multi-snapshot sensing on the overall performance. 
This work extends the \ac{AE} model of~\cite{MateosRamos2021} by adding multiple target capabilities for detection and localization.
%% description of hypothesis, methods, objectives
We describe the detection of multiple targets not as a classification task with the number of targets as classes but instead design it as parallel detection tasks resulting in the novel counting encoding. 
%We show its effectiveness in simulations. Furthermore, we are faced with sets of values in our simulations. We show different methods of how this extension to sets can be handled without adding a lot of system complexity.
The permutation invariance of targets during detection brings additional challenges to the training of the \acp{NN}. To address this issue, we present multiple approaches with low additional complexity.

%%% Paragraph 5: Roadmap paper
%The remainder of this paper is structured as follows:    
\section{System Model}
\begin{figure*}
	\centerline{\includegraphics{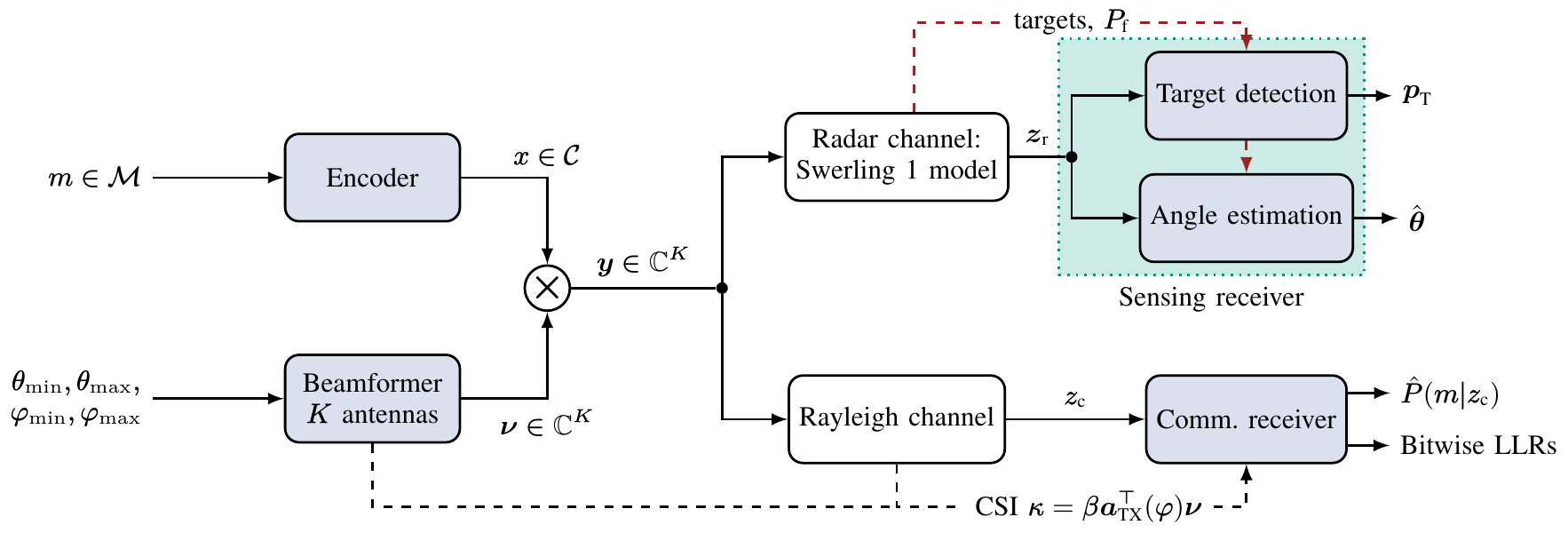}}
	\vspace{-0.5cm}
	\caption{\ac{JCAS} autoencoder as proposed in \cite{MateosRamos2021}, light blue blocks are trainable \acp{NN}, red dashed paths are only active while propagating the training data}
	\label{fig:flowgraphtrain}
	\vspace*{-0.2cm} 
\end{figure*}
The system block diagram, shown in Fig. \ref{fig:flowgraphtrain}, is based on \cite{MateosRamos2021}.
The encoder transforms the data symbols $m\nobreak \in \nobreak\mathcal{M}:=\nobreak \{1,2,\ldots,M\}$ into complex modulation symbols $x \in \mathcal{C} \subset \mathbb{C}$, with $|\mathcal{C}|=M$. The complex symbols are multiplied with a unique $\nu_i = g_i\exp(\j\gamma_i)$ for each antenna $i$ with beamforming gain $g_i$ and phase shift $\gamma_i$ to steer the signal to our areas of interest. The encoder and beamformer employ power normalization to fulfill power constraints. We consider a maximum of $T_{\max}$ radar targets and a linear array of $K$ antennas in the transmitter and the radar receiver. The beamformer inputs are the azimuth angle regions in which communication and sensing should take place. The communication receiver is situated randomly in the interval $[\varphi_{\min},\varphi_{\max}]$ and the radar target positions are uniformly drawn from $[\theta_{\min},\theta_{\max}]$.
The transmit signal $\vect{y}$ is fed into a Rayleigh channel before being received by the communication receiver with a single antenna as
\begin{align}
	z_{\text{c}} =  \beta \vect{a}_{\text{TX}}(\varphi)^\top \vect{y} + n,
\end{align}
with complex normal distributed $\beta \sim \mathcal{CN}(0,\sigma_\text{c}^2)$ and $n\nobreak \sim \nobreak \mathcal{CN}(0,\sigma_\text{n}^2)$.
We assume that channel estimation has already been performed, therefore the \ac{CSI} $\kappa=\beta \vect{a}_{\text{TX}}^\top(\varphi)\vect{\nu}$ is available at the communication receiver. The input of the communication receiver is ${z}_{\text{c}}/\kappa$. The outputs of the receiver are estimates of the symbol-wise maximum a posteriori probabilities that are transformed into bitwise \acp{LLR} that can be used as input to a soft-decision channel decoder. 

For the simulation of multiple radar targets, we express the sensing signal that is reflected from $T$ radar targets as
\begin{align}
	\vect{z}_{\text{r}} = \left(\sum_{k=0}^{T} \alpha_k \vect{a}_{\text{RX}}(\theta_k) \vect{a}_{\text{TX}}(\theta_k)^\top \vect{y} \right)+ \vect{n},
\end{align}
with the radar targets following independently a Swerling-1 model $\alpha_k \sim \mathcal{CN}(0,\sigma_{\text{r}}^2)$ and $\vect{n} \sim \mathcal{CN}(0,\sigma_{\text{n}}^2 \mat{I})$. The signal propagation from $K$ antennas toward an azimuth angle $\theta_k$ is modeled with the spatial angle vector $\vect{a}_{\text{TX}}(\theta_k) \in \mathbb{C}^K$ whose entries are given by
\begin{align}
	[\vect{a}_{\text{RX}}(\theta_k)]_{i} = [\vect{a}_{\text{TX}}(\theta_k)]_i = \exp\left(\j2\pi  \left(\frac{d_y}{\lambda} i  \sin\theta_k\right)\right).
\end{align}
The parameter $d_y$ describes the horizontal distance between each antenna element at the transmitter and the radar receiver.
Target detection and angle estimation are both performed using $\vect{z}_{\text{r}}$. The output of the target detection \ac{NN} is a probability vector $\vect{p}_{\text{T}} \in [0,1]^{T_{\max}}$. Each entry of $\vect{p}_{\text{T}}$ denotes the probability that a specific target is present, without a specific order. From $\vect{p}_{\text{T}}$, we determine the number of detected targets.
The angle estimation block outputs a vector $\hat{\vect{\theta}} \in [- \frac{\pi}{2}, \frac{\pi}{2}]^{T_{\max}}$ denoting the estimated azimuth angle of each target.

With a Swerling-1 model, we model scan-to-scan deviations of the \ac{RCS}. During training of target detection, the values $\alpha_k$ remain equal over all receive antennas, while being independently sampled from the complex normal distribution for different targets or different time instants. 

Our system is designed to solve three different tasks:
\begin{itemize}
	\item transmit data over a Rayleigh channel,
	\item estimate the number of targets in our angle region of interest (detection),
	\item estimate the position of the targets (angles of arrival).
\end{itemize}
Considering a possible upsampling with $u>1$, we combine outputs of the sensing receiver by averaging the detection probabilities along the upsampling axis. Similarly, we average the estimated angles after having applied the corresponding set method discussed in Sec. \ref{sec:setmethods}.

\subsection{Angle Estimation Benchmark}
\acused{ESPRIT}
We use the well-studied \ac{ESPRIT} algorithm as a benchmark for angle estimation as studied in \cite{Trees2002, Yilmazer2010}.
%This subspace algorithm estimates the angles $\theta_i$ by:
%\begin{enumerate}
%	\item Calculating the eigendecomposition $\mat{U}$ of $\mat{C} = \vect{z}_r \vect{z}^H_r$
%	\item Slicing $\mat{U}$ into two parts by discarding the first column for $\mat{S}_1$ and the last column for $\mat{S}_2$
%	\item Calculating the eigenvalues of $\Theta = [\mat{S}_1^H \mat{S}_1]^{-1}\mat{S}_1^H \mat{S}_2$
%	\item From the $t$ largest eigenvalues $\lambda_i$ estimate the angles as $\hat{\theta_i}=\sin^{-1}\left(\frac{\arg(\lambda_i)}{\pi}\right)$
%\end{enumerate}
%
The estimation variance of ESPRIT increases when the number of snapshots is small, therefore we also adapt \ac{ESPRIT} for single snapshot evaluation as described in \cite{WLWLAF2020}, by constructing a Hankel matrix before auto-correlation to improve the estimation \ac{RMSE}.
%Additionally, \ac{CRB} is usually not a tight bound for the ML estimator when estimating from a small number of snapshots, therefore we do not consider it as a useful bound for these single snapshot estimations \cite[Chapter 8.2]{Trees2002}.
For validation purposes, 
we only measure the \ac{RMSE} for all targets that were detected by the target detection block and are also present. We assume that in cases where the target detector fails at recognizing a target, the reflected signal power from the target is very low or there is another target extremely close to it and its reflection is shadowed. Therefore calculating the error only for detected targets can lead to a higher effective \ac{SNR} by ignoring low power reflections in the evaluated samples.

%\ac{ESPRIT} grows linearly in complexity with a growing number of samples, just as the \ac{NN} approach considering we propagate each sample through the \acp{NN}.
%\ac{ESPRIT} has a complexity of
%\begin{align}
%	\mathcal{O}\{4K^2N+8K^6-8KT^2+8K^2T^2+9T^3\}
%\end{align}
%flops \cite{Li2018}. In comparison, the \ac{NN} forward pass complexity arises to 
%\begin{align}
%	\mathcal{O}\{ N (4K^2+4K^2+2K^2+KT_{\max}) \},
%\end{align}
%when assuming the use of the rectified linear unit as an activation function after training (which takes 1 flop). XXX
%high complexity compared with performing a forward pass in the angle estimation \ac{NN}.

%\section{Neural Network Configurations}
\subsection{Neural Network Training and Validation}\label{sec:NNconfig}
We realize all blocks in transmitter and receiver highlighted in Fig.~\ref{fig:flowgraphtrain} by \acp{NN}, which are jointly trained in an end-to-end manner.
We utilize fully connected \ac{NN} layers with an \ac{ELU} activation function. The number of neurons and the output functions vary according to the task and are summarized in Tab.~\ref{tab:NNstructure}. Although arriving at a similar structure to \cite{MateosRamos2021}, we couple the \ac{NN} layer size with different system parameters.
The fully connected \acp{NN} each of depth 5 have different layer widths; each list item denotes the number of neurons in a layer of the \ac{NN}. The output layer size of encoder and beamformer requires two neurons to represent each complex output value with two real numbers. Consequently, the number of input neurons for target detection, angle estimation, and the communication receiver also use two real-valued inputs to represent complex input signals. The encoder and beamformer are subject to power normalization representative for the power constraints of a radio transmitter. The decoder output uses a softmax layer to generate probabilities $\hat{P}(m|z_{\text{c}})$. 
We set the learning rate to $0.001$ for all \acp{NN} and employ the Adam optimizer. We use $20\cdot T_{\max}$ mini-batches with $N_\text{mb}=10^4$ samples in each epoch and train for $150$ epochs, resulting in convergence of the \ac{NN} training.
\begin{table}
	\centering
	\caption{Structure of \acp{NN} of the \ac{JCAS} system}
	\begin{tabular}{r l l}
		\toprule
		Subnet & Network structure & Output layer \\
		\midrule
		Encoder & $[M,2M,2M,2M,2]$ & mean power norm \\
		%\hline
		Beamformer & $[5,K,K,2K,2K]$ & power norm \\
		%\hline
		Decoder & $[2,2M,2M,2M,M]$ & softmax \\
		%\hline
		Target detection& $[2K,2K,2K,K,T_{\max}]$ & sigmoid \\
		%\hline
		Target angle estimation& $[2K,2K,2K,K,T_{\max}]$ & $\frac{\pi}{2}\cdot$tanh$(\cdot)$ \\
		\bottomrule
	\end{tabular}
	\label{tab:NNstructure}
	\vspace*{-0.2cm}
\end{table}

During training, additional knowledge is injected into the \acp{NN} as shown in Fig. \ref{fig:flowgraphtrain}.
To decouple both sensing tasks during training, the actual number of radar targets is injected into the angle estimation network by only propagating through the \ac{NN} if one or more targets are present.
During validation, we measure the \ac{BMI} of the communication receiver. Since the \ac{JCAS} system learns both symbol constellation and bitmapping, this is the most suitable metric \cite{Cammerer2020}.

\subsection{Loss Functions}
We need a combined loss function to jointly optimize our different networks.
\begin{enumerate}
	\item Communication Loss: 
	As proposed in \cite{Cammerer2020}, we use the \ac{BCE} as a loss function $L_{\text{comm}}$ to optimize mainly the encoder, decoder, and beamformer. Since this loss function takes the \ac{BMI} into account, the complex symbol alphabet and the bit mapping are jointly optimized.
	\item Detection Loss:
	We utilize the \ac{BCE} between estimated and present targets as a loss function $L_{\text{detect}}$. This optimization mainly affects the target detection and the beamformer.
	\item Angle Estimation Loss: 
	We use a \ac{MSE} loss between valid and estimated angles as a loss function $L_{\text{angle}}$, which mainly affects angle estimation and the beamformer.
\end{enumerate}
We propose a training schedule consisting of three different training stages to improve the results. Therefore we adapt the loss function after a third and two-thirds of all training epochs.
Different loss terms are weighted and added to enable joint training. 
The loss functions $L_i$ of the different training stages are: 
\begin{align}
	L_1 &= (1-w_{\text{r}})\cdot L_{\text{comm}}  + w_{\text{r}} w_{\text{a}} \cdot L_{\text{angle}},\\
	L_2 &= (1-w_{\text{r}})\cdot L_{\text{comm}} + w_{\text{r}} \cdot L_{\text{detect}},\\ 
	L_3 &= (1-w_{\text{r}})\cdot L_{\text{comm}} + w_{\text{r}} \cdot L_{\text{detect}} + w_{\text{r}} w_{\text{a}} \cdot L_{\text{angle}}.
\end{align}
We choose a weighting factor of $w_{\text{r}}=0.9$. 
Since both communication and sensing functionalities profit from a high \ac{SNR}, the beamformer is trained to radiate most energy toward the possible positions of communication receiver and radar target. Since only limited power is available, $w_{\text{r}}$ affects the magnitude of the beam in direction of the radar targets and the direction of the communication receiver by being able to change the optimal power trade-off of communication and sensing.
By increasing $w_{\text{r}}$, we can increase the importance of the sensing functionality, therefore increasing the radiated power towards $[\theta_{\min},\theta_{\max}]$ but decreasing the radiated power toward the communication receiver in $[\varphi_{\min},\varphi_{\max}]$. The other weighting factor was chosen to $w_{\text{a}}=20$ to further improve the angle estimation.

The training schedule has the effect that initially everything but the target detection is trained. The effect of the angle estimation on the transmit beam is comparably weak; this leads to a good initial performance of the communication part while the angle estimation is trained to extract features from reflections with comparably low power. 
Afterwards, switching the angle estimation with target detection in $L_2$ results in a beamform radiating mostly toward our angle ranges of interest while $w_{\text{r}}$ controls the ratio of average radiated power in $[\theta_{\min},\theta_{\max}]$ and $[\varphi_{\min},\varphi_{\max}]$.
Lastly, applying the fully joint loss function $L_3$ accelerates the training of the angle estimation as well as target detection, when the communication part has almost converged.

\subsection{One-hot vs. Counting Encoding}\label{sec:encoding}
To extend the system from the one target case as proposed in \cite{MateosRamos2021}, we need to decide how to encode different numbers of detectable targets. To model partially correct detection, e.g., detection of one target when two are present, we propose a novel representation called \emph{counting encoding} that can be understood as a subcategory of multi-hot encoding. It enables direct measurement of detection probabilities and notably supports choosing a resulting false alarm rate. In essence, the detection of $T_{n}$ targets gets divided into $T_{\max}$ tasks to confirm the presence of a maximum of $T_{\max}$ targets.
The detection vector $\vect{c}$ that represents $T_i$ targets is built with
\begin{align}
	c_{i} = \begin{cases}
		1 & \text{if }i \leq T_i,\\
		0 & \text{otherwise},
	\end{cases} \quad \text{for } i= 1,\ldots,T_{\max}.
\end{align}
For an example with $T_{\max}=3$, the encoded vectors $[0,0,0],[1,1,1]$ and $[1,1,0]$ represent the occurrence of zero, three, and two targets. By summation, we can recover the number of targets and by element-wise multiplication with the angle estimates, we can mask the angle estimate vectors $\hat{\vect{\theta}}$ to match the number of targets present.
We can train the target detection \ac{NN} with a sigmoid output layer and transform the logits $\ell_n$ into probabilities $c_{\text{est},n} = \nobreak \sigma(\ell_n)$ with
\begin{align}
	c_{\text{est},n} = P(\text{``}n\text{ or more targets detected''}).
\end{align}
We introduce a weighted false alarm rate that emphasizes the number of targets falsely detected. Counting encoding implicitly supports this weighting when summing over multiple entries since the event described by $c_{\text{est},n}$ includes $c_{\text{est},n+1}$.
We calculate both the detection rate $P_{\text{d}}$ and the weighted false alarm rate $P_{\text{f}}$ from the valid $T_n$ targets in timestep $0 \leq n \leq N-1$ with $\mat{C} \nobreak \in \nobreak \{0,1\}^{N\times T_{\max}}$ and the estimated targets $\mat{C}_{\text{est}} \in \nobreak [0,1]^{N\times T_{\max}}$ as
\begin{align}
	P_{\text{d}} &= \frac{1}{\sum_{n=0}^{N-1} T_n} \sum_{i=1}^{N} \sum_{j=1}^{T_n} \round{c_{\text{est},i,j}},\\
\intertext{and}
	P_{\text{f}} &= \frac{1}{\sum_{n=0}^{N-1} (T_{\max}-T_n)} \sum_{i=1}^{N} \sum_{j=T_n+1}^{T_{\max}} \round{c_{\text{est},i,j}},
\end{align}
where $\round{\cdot}$ denotes rounding to the next integer.
%\todo{Rephrase to be more clear on how Pf is controlled}
During validation, the target detection probability is sorted in descending order. This ensures $c_{\text{est},n+1} \leq c_{\text{est},n}$. The detection output remains therefore easily interpretable by preventing impossible states, e.g., no detection of a first target but still detection of a second target. 
This sorting is arguably necessary to interpret all possible outputs, but it should be already performed by the detection \ac{NN} since we do not sort during training.

Since traditional one-hot encoding is prevalently in use for classification problems as in \cite{OShea2017}, we adapt the target detection \ac{NN} for one-hot encoding as a benchmark alternative. We add one neuron to the output layer and replace the sigmoid function with softmax.
%We do not consider modeling target detection as a regression problem, since we have no approach to target a weighted false alarm rate in that model. 
We denote the valid one-hot matrix as $\mat{O} \in \nobreak \{0,1\}^{N\times (T_{\max}+1)}$ and the estimated targets as $\mat{O}_{\text{est}}\nobreak \in \nobreak [0,1]^{N\times (T_{\max}+1)}$ describing the presence of $0,1,\ldots,T_{\max}$ targets.
For the one-hot encoding, detection probability and the weighted false alarm rate are calculated using the hard-decision $h_n=\arg\max_k (o_{\text{est},n,k})$ as
\begin{align}
	P_{\text{d,onehot}} &= \frac{\sum_{n=0}^{N-1} %\sum_{k=1}^{T_n} k \cdot o_{\text{est},n,k}
		%\arg\max_k (o_{\text{est},n,k})
	%	\sum_{k=0}^{\min\{T_n,h_n\}} k}{\sum_{n=0}^{N-1} T_n} 
	\min\{T_n,h_n\}}{\sum_{n=0}^{N-1} T_n} 
	,
\intertext{and}
	%P_{\text{f,onehot}} &= \frac 1N \sum_{n=0}^{N-1} \sum_{k=1}^{T(n)} o_{\text{est}}(n,k) \frac{n}{T(n)+1} \sum_{m=0}^{k-1} o_{\text{valid}}(n,m)
	P_{\text{f,onehot}} &=  \frac{\sum_{n=0}^{N-1} 
	%\sum_{k=T_n}^{\max\{T_n,h_n\}} (k-T_n) %\cdot 
	(\max\{T_n,h_n\}-T_n)
	%o_{\text{est},n,k} 
	%\frac 12 \left(|\arg\max_k (o_{\text{est},n,k})-T_n|+(\arg\max_k (o_{\text{est},n,k})-T_n)\right)
	%\frac 12 \left( |h_n-T_n|+(h_n-T_n)\right)
}{\sum_{n=0}^{N-1} (T_{\max}-T_n)}.
\end{align}
The probability vectors can be transformed from one-hot encoding to counting encoding by
\begin{align}
	c_{\text{est},k} &= \sum_{n=k}^{T_{\max}}  o_{\text{est},n}, \\
	\intertext{and for counting encoding to one-hot encoding using}
	o_{\text{est},k} &= \begin{cases}
		c_{\text{est},k} & \text{ for } k=T_{\max},\\
		c_{\text{est},k} - c_{\text{est},k+1} & \text{ for } k \in [1,T_{\max}-1],\\
		%1-\nobreak\sum_{k=1}^{T_{\max}}o_{\text{est}}(k) & \text{, for } k=0.
		1-c_{\text{est},1} & \text{ for } k=0.
	\end{cases} 
\end{align}
\subsection{Fixed False Alarm Rate}
For many applications, the implications of a false alarm and a missed detection are different. For example in automotive driving or malicious drone detection, the actions associated with detection and non-detection are so vastly different that the probability of false alarm and missed detection should be different.
We train for a fixed weighted false alarm rate (meaning the probability that a target is detected even though none are present), but our model can easily be adapted to train for a fixed missed detection rate.
During training, we proceed as follows:
\begin{itemize}
	\item choose all output logits $\ell_n$ of the target detection with $c_{n}=0$, $n \in [0,N-1]$, with $X = \sum_{n=0}^{N-1}T_n$ being the number of chosen logits in the whole training minibatch,
	\item sort these logits in ascending order,
	\item choose $\ell_i$ with $i = \lfloor (1-P_{\text{f}})\cdot X\rfloor$,
	\item subtract $\ell_i$ from all logits and set $\ell_{\text{off}}=\ell_i$, and
	\item apply the sigmoid function $c_{\text{est},n} = \sigma(\ell_n)$.
\end{itemize}
During validation, we set $c_{\text{est},n} = \sigma(\ell_n-\ell_{\text{off}})$ without updating $\ell_{\text{off}}$, ensuring the same system behavior during validation. For multiple target detection, one $\ell_{\text{off}}$ is used for $\mat{C}_{\text{est}}$.

In order to specify a targeted $P_{\text{f}}$ using one-hot encoding, 
%we introduce an additional \ac{MSE} loss of targeted and achieved $P_{\text{f}}$. This loss term needs to converge to $0$ to achieve the targeted weighted false alarm rate. Alternatively, 
we offset the output probabilities of the \ac{NN} with $P_{\text{off}} = \nobreak (P_{\text{f}}-\nobreak P_{\text{f,onehot}})\cdot [1,-\frac 1T_{\max},-\frac 1T_{\max},\ldots,-\frac 1T_{\max}]^\top$ after calculating the resulting weighted false alarm rate $P_{\text{f,onehot}}$ (without using hard-decision to improve training stability). To ensure probability values in $[0,1]$, we clip at these extremes.
%For our final results, offsetting the output probabilities achieved a weighted false alarm rate closer to $P_{\text{f}}$.
Using one-hot encoding, we replace the binary cross-entropy loss for target detection with the cross-entropy loss, handling the optimization as a classification problem.

\subsection{Sequence Ambiguity in Multiple Target Detection}\label{sec:setmethods}
For simulation purposes, we face the fact that real and estimated angles exist as vectors in our system, while we need to compare distances of sets. The order in which our \ac{NN} estimates the angles of different targets is practically not important, but we need to be able to match estimates to their valid counterpart. We have multiple approaches to handle this extension to sets during training of the \ac{NN}.

\subsubsection{Sortinput} This simple approach sorts all input angles in our validation set. This corresponds to an additional task to the angle estimation \ac{NN}: Not only estimating the correct angles but also returning them in order. This approach is effective if the angles are estimated correctly.

\subsubsection{Sortall} This extension of the first approach sorts the validation set and the outputs of the \ac{NN}. If angle estimations are correct, this set behavior represents a translation to vectors. 
%But since the \ac{NN} can return faulty angles, sorting could also worsen the measured \ac{MSE}. 
We expect the sortall approach to perform at least as well as sortinput.

\subsubsection{Permute} For this method, the angle permutation that minimizes the \ac{MSE} is chosen as the correct permutation, and returned vectors are permuted according to it. This represents the best possible method concerning \ac{MSE} but brings also significant overhead since $T!$ angle permutations need to be considered.

We calculate the average complexity for one sample for the different set approaches, shown in Tab. \ref{tab1}. For sortinput and sortall, we assume a Quicksort algorithm.
\begin{table}[ptb]
	\caption{Complexity of different Set methods with $T$ different targets}
	\begin{center}
		\begin{tabular}{c c c c}
			\toprule
			\textbf{Method} & Sortinput & Sortall & Permute \\
			\midrule
			\textbf{Complexity} & $ \mathcal{O}(T \log (T))$ & $\mathcal{O}(2T \log (T))$ & $\mathcal{O}(T!)$\\ %\frac 12 T!)$ \\
			\bottomrule
		\end{tabular}
		\label{tab1}
	\end{center}
	\vspace*{-0.2cm}
\end{table}
%Further simulations will show whether the sorting approach is sufficient for the handling of the occurrence of sets. 
If the \ac{NN} estimation in one of the sorting approaches contains angle estimates far away from the true angle, the overall \ac{MSE} could be much larger than expected as the whole sorting is faulty. For example, if $\hat{\theta}_k > \hat{\theta}_{k+1}$ but $\theta_k < \theta_{k+1}$, the values are switched for evaluation even if $\hat{\theta}_k\approx {\theta}_k$.
During validation, we use the permute method for all trained \acp{NN}.

\section{Simulation Results}
In our simulations, the communication receiver is situated at an \ac{AoA} of $\varphi \in [30^\circ,50^\circ]$. The radar targets are found in $\theta \in [-20^\circ,20^\circ]$.
Our monostatic sender and radar receiver are simulated as a linear array with 16 antennas. For the radar receiver, we target a weighted false alarm rate of $P_{\text{f}}=\nobreak10^{-2}$ while optimizing the detection rate and the angle estimator.

\subsection{Communication Results}
Previous works \cite{Cammerer2020, OShea2017} have shown that an \ac{AE} approach to substitute modulation and demodulation is effective. In combination with sensing, constellation diagrams tend to assume a PSK-like form. This behavior can be explained intuitively, since sensing profits greatly from a constant signal amplitude. For $M=8$ and a communication \ac{SNR} of $\sigma_{\text{c}}^2 / \sigma_{\text{n}}^2 \widehat{=}20\,$dB,
we achieve a \ac{BMI} of up to $2.94\,$bits that enables effective communication. The beamformer achieves an average gain of $2.7\,$dB in the angle range of the communication receiver. For the single target results, we use $M=4$ to have comparable results to \cite{MateosRamos2021}, achieving a \ac{BMI} of $1.87\,$bits with a beamformer gain toward the communication receiver of $-4\,$dB. We can see that for $M=4$ we lowered our channel \ac{SNR} but we simultaneously improved the \ac{SNR} of our sensing channel.
%The channel capacity can be calculated according to \cite{Alouini1999}, resulting in $\frac{C}{B}=4.6\,$bit/s/Hz for the $M=4$ case and $\frac{C}{B}=6.7\,$bit/s/Hz for $M=8$.

\subsection{Single Target Results}
\begin{figure}
	\includegraphics{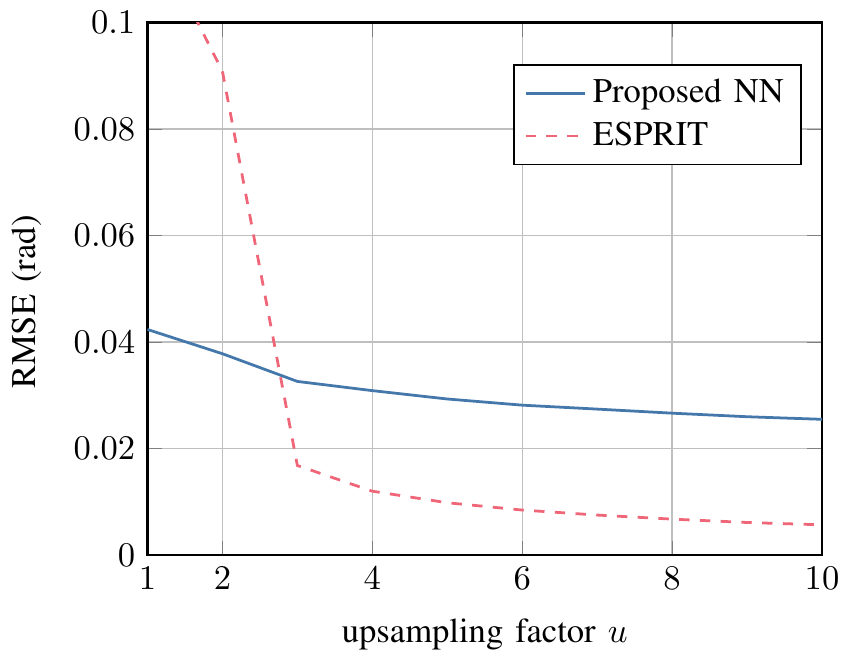}
	%\tikzsetnextfilename{JCaS-canc}
	% \begin{tikzpicture}
	% 	\begin{axis}[
	% 		yscale=0.95,
	% 		xlabel=upsampling factor $u$, ylabel=RMSE (rad),
	% 		grid=major,
	% 		legend entries={Proposed NN, ESPRIT}, 
	% 		legend cell align={left},
	% 		legend pos=north east,
	% 		xmin=1,xmax=10,
	% 		ymin=0,ymax=0.1,
	% 		axis line style=thick,
	% 		extra x ticks={1},
	% 		extra x tick style={grid=none,tickwidth=0},
	% 		tick label style={/pgf/number format/fixed},
	% 		]
	% 		\addplot[mark=none, color=cb-1, thick] table [x = cpr, y = nofft]
	% 		{\figures/rmse_vs_cpr_1targ.txt};
	% 		\addplot[mark=none, color=cb-2, dashed, thick] table [x = cpr, y = ESPRIT]
	% 		{\figures/rmse_vs_cpr_1targ.txt};
	% 	\end{axis}
	% \end{tikzpicture}
	\vspace{-0.2cm}
	\caption{\ac{RMSE} for present and detected targets with and without cancellation with ESPRIT benchmark \ac{SNR} of $0\,$dB for the radar channel and $20\,$dB for the communication channel ($M=4$) and $1$ possible target}
	\label{fig:rmse_cpr_canc}
	\vspace{-0.2cm}
\end{figure}

Results with a single target were already presented in \cite{MateosRamos2021}.
In this work, we introduce a different benchmark. In the single snapshot case (upsampling factor $u=1$) and for $u=2$, the proposed system outperforms the \ac{ESPRIT} algorithm. When considering multiple snapshots with $u \geq 3$, ESPRIT outperforms the angle estimation \ac{NN}, as can be seen in Fig.~\ref{fig:rmse_cpr_canc}. The simple approach of taking the mean of the \ac{NN} output when increasing the number of samples seems to be inferior to using the covariance estimate based on all recorded samples.

%The trained angle estimation applies averaging over $u$ estimated angles. 
\begin{figure}
	\includegraphics{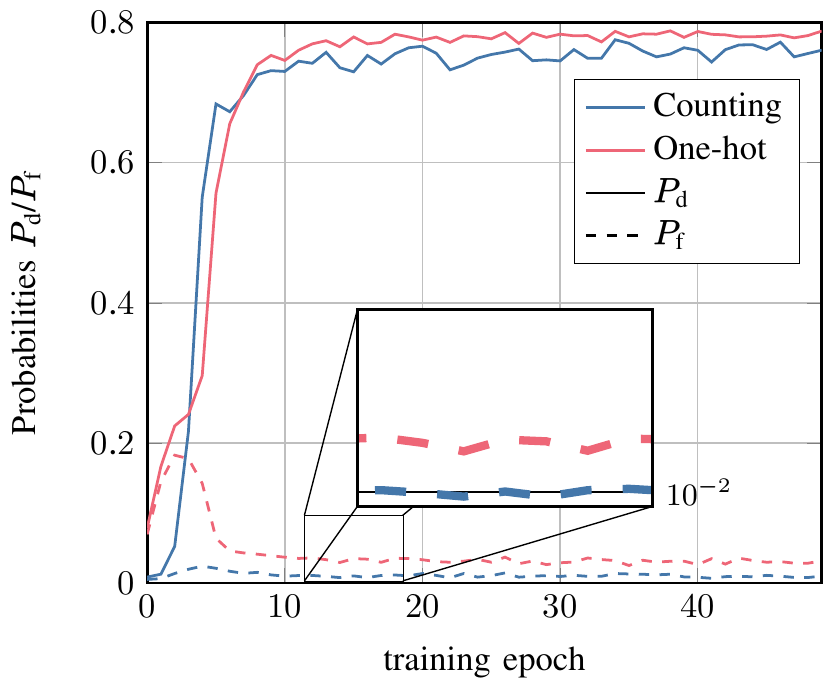}
	\vspace{-0.2cm}
	\caption{Comparison of detection and weighted false alarm rate of one-hot and counting encoding for an \ac{SNR} of $0\,$dB for the radar channel and $20\,$dB for the communication channel ($M=4$) for $1$ radar target}
	\label{fig:encoding-1targ}
	\vspace{-0.2cm}
\end{figure}

Comparison of counting encoding and one-hot encoding shows their suitability for the studied problem, yet control of the weighted false alarm rate is much tighter in the counting encoding as shown in Fig. \ref{fig:encoding-1targ}. We choose the same training parameters for both simulations, with an evaluation of $20$ batches of $N=10^4$ values for each training epoch.
%The encoding itself has a negligible effect on the other training metrics as angle estimation and communication. 
During training, the weighted false alarm rate of the counting encoding never exceeds the targeted value $P_{\text{f}}$ by more than $10\%$. Meanwhile, the one-hot encoding oscillates around a weighted false alarm rate of $P_{\text{f}}\approx 3 \cdot 10^{-2}$.
%Introducing an \ac{MSE} loss between targeted value and $P_{\text{f}}$ performed even worse than modifying the detection probabilities as described in Sec. \ref{sec:encoding}.
For applications that generally need to ensure that a given weighted false alarm rate is kept, the counting encoding is more promising. Additionally, counting encoding has computational advantages: The target detection \ac{NN} output layer saves one neuron and the estimated angle vector can be directly element-wise multiplied with the decision output to calculate one angle estimate for each detected target. 

\subsection{Multiple Target Results}
Next, we consider the detection of multiple targets with $T_{\max}=3$ while keeping the communication SNR and radar SNR both at $20\,$dB. We trained the system with a total of $9\cdot 10^7$ samples.
Training angle estimation and target detection sequentially followed by joint training decreased the angle \ac{RMSE} from roughly~$0.1$ to $0.04$. By repeating each training epoch for $1$ to $T_{\max}$ targets, the detection rate for one snapshot rose from approximately $0.6$ to $0.8$.
For the radar path, the introduction of multiple targets means that we now observe multiple reflections, leading to an increased \ac{SINR}.
Comparison of the results of the two encoding metrics shows again how the counting encoding stabilizes the weighted false alarm rate around $P_{\text{f}}\approx 0.01$, while the one-hot encoding settles at $P_{\text{f}}\approx 0.0002$. This causes a much lower detection rate of approximately $0.5$. We also reach a higher \ac{RMSE} of $0.06$ for the estimated angles.

\begin{figure}
	\includegraphics{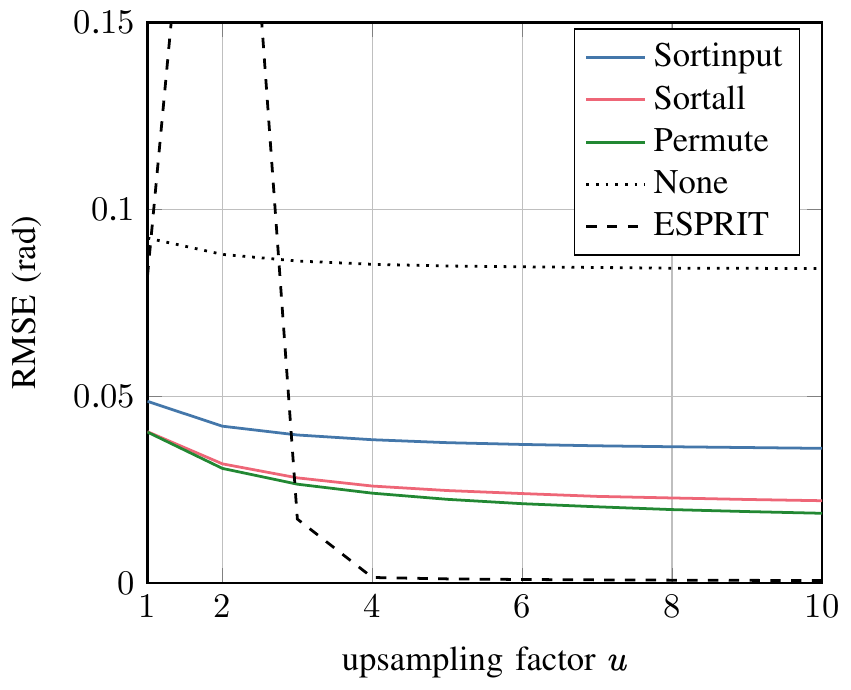}
	% \tikzsetnextfilename{JCaS-3targs}
	% \begin{tikzpicture}
	% 	\begin{axis}[xlabel=upsampling factor $u$, ylabel=RMSE (rad),
	% 		grid=major,
	% 		legend entries={Sortinput, Sortall, Permute, None, ESPRIT},
	% 		legend pos=north east,
	% 		xmin=1,xmax=10,
	% 		ymin=0,ymax=0.15,
	% 		tick label style={/pgf/number format/fixed},
	% 		axis line style=thick,
	% 		extra x ticks={1},
	% 		legend cell align={left},
	% 		legend style={
	% 			at={(0.8,0.99)},
	% 			anchor=north,
	% 		},
	% 		extra x tick style={grid=none,tickwidth=0},
	% 		%yscale=0.9,
	% 		]
	% 		\addplot[mark=none, color=cb-1,thick] table [x = cpr, y = sortphi]
	% 		{\figures/rmse_vs_cpr_3targs_noset.txt};
	% 		\addplot[mark=none, color=cb-2,thick] table [x = cpr, y = sortall]
	% 		{\figures/rmse_vs_cpr_3targs_noset.txt};
	% 		\addplot[mark=none, color=cb-3, thick] table [x = cpr, y = permute]
	% 		{\figures/rmse_vs_cpr_3targs_noset.txt};
	% 		\addplot[mark=none, color=black, dotted, thick] table [x = cpr, y = none]
	% 		{\figures/rmse_vs_cpr_3targs_noset.txt};
	% 		\addplot[mark=none, color=black, dashed, thick] table [x = cpr, y = ESPRIT]
	% 		{\figures/rmse_vs_cpr_3targs_noset.txt};
	% 	\end{axis}
	% \end{tikzpicture}
	\vspace{-0.2cm}
	\caption{\ac{RMSE} for present targets with ESPRIT benchmark \ac{SNR} of $\sigma_{\text{r}}^2 / \sigma_{\text{n}}^2 =\nobreak 20\,$dB for the radar channel and $20\,$dB for the communication channel ($M=\nobreak8$, $T_{\max}=3$)}
	\label{fig:rmse_cpr_3targs}
	\vspace{-0.2cm}
\end{figure}

In Fig. \ref{fig:rmse_cpr_3targs}, we plot the \ac{RMSE} of the angle estimation for detected and present targets for $10^5$ transmissions versus the upsampling factor $u$. We compare the different set methods from Sec. \ref{sec:setmethods} and also show the ESPRIT benchmark.
For the multiple target case, the different set methods enable training of the \acp{NN}. The method labeled ``None'' denotes \ac{NN} training without using any set method and shows that the implementation of a set method for multiple target estimation is necessary. The permute method performs the best, which was expected since it considers all possible set permutations while still using the \ac{MSE} loss. The methods based on sorting perform relatively well and are only slightly outperformed by permuting. 
These set methods outperform the ESPRIT benchmark for small upsampling factors $u \leq 3$. The specific single-snapshot ESPRIT implementation as used for $u=1$ cannot outperform the proposed system.

The detection probability is comparable for all set methods. The weighted false alarm rates are also similar for all methods and converge from the targeted $P_{\text{f}}$ to zero with an increasing $u$.
The detection rate saturates to a value of $0.83$ while increasing $u$. For increased detection rate for rising $u$, the detection threshold needs to be further modified. 

\section{Conclusion}
\acresetall
\acused{ESPRIT}
In this work, we demonstrate the feasibility of the \ac{AE} approach to \ac{JCAS} for multiple targets.
We evaluated different set methods that enable training of angle estimation for multiple targets. Depending on the permissible system complexity, all three options remain contenders for application in future systems.
We outperformed an \ac{ESPRIT} benchmark for angle estimation for small upsampling factors $u$. 
The novel counting encoding enables setting a design false alarm rate that constraints the detection rate of a \ac{NN} target detector. 
We see counting encoding as a promising alternative to classification using one-hot encoding for problems that include object recognition connected with counting. 
The proposed method is particularly suitable for \ac{JCAS} systems, where the number of available snapshots is typically limited.

\footnote[]{The simulation code is available at \url{https://github.com/frozenhairdryer/JCAS_multitarg}}

\bibliography{literature_short.bib}
\bibliographystyle{IEEEtran}

\end{document}